\documentclass[aps,prd,twocolumn,superscriptaddress]{revtex4}
\usepackage{graphicx}
\usepackage{amsmath}
\usepackage{epsfig}
\usepackage{amsmath,amsfonts,amssymb}
\usepackage{times}
\usepackage{epsfig}
\usepackage{epstopdf}
\usepackage[
colorlinks=true,
linkcolor=blue,      
citecolor=blue,      
urlcolor=blue        
]{hyperref}

\usepackage{subcaption}
\usepackage[export]{adjustbox}

\usepackage{graphicx}	
\usepackage{amsmath}	
\usepackage{epsfig}
\usepackage{epstopdf}
\newcommand{\Sfr}{\mathfrak{S}}
\newcommand{\rhobar}{\bar{\rho}}
\newcommand{\Pbar}{\bar{P}}
\newcommand{\wbar}{\bar{w}}

\let\origss\ss
\renewcommand{\ss}{\mbox{\origss}}   

\def\be{\begin{equation}}
	\def\ee{\end{equation}}

\usepackage{color}

\begin{document}
\baselineskip=12pt
\def\black{\textcolor{black}}
\def\red{\textcolor{black}}
\def\blue{\textcolor{blue}}
\def\green{\textcolor{black}}
\def\be{\begin{equation}}
\def\ee{\end{equation}}
\def\bea{\begin{eqnarray}}
\def\eea{\end{eqnarray}}
\def\orc{\Omega_{r_c}}
\def\om{\Omega_{\text{m}}}
\def\E{{\rm e}}
\def\bearst{\begin{eqnarray*}}
\def\eearst{\end{eqnarray*}}
\def\peleven{\parbox{11cm}}
\def\peffec{\peight{\bearst\eearst}\hfill\peleven}
\def\pspace{\peight{\bearst\eearst}\hfill}
\def\ptwelve{\parbox{12cm}}
\def\peight{\parbox{8mm}}

\newcommand{\nk}[1]{\textcolor{red}{[{\bf Nima}: #1]}}

\newcommand{\hm}[1]{\textcolor{blue}{[{\bf Hossein}: #1]}}


\title{ Gravitational-Susceptibility Dip Behind DESI’s results:\\ Addressing the Phantom Crossing by Local limit of Nonlocal Gravity}

\author{Aria Hemmatian}
\address{Department of Physics, Sharif University of
	Technology, P.~O.~Box 11155-9161, Tehran, Iran}
\author{Yahya Mohammadi}
\address{Department of Physics, Sharif University of
	Technology, P.~O.~Box 11155-9161, Tehran, Iran}
\author{Javad Tabatabaei}
\address{Department of Physics, Sharif University of
	Technology, P.~O.~Box 11155-9161, Tehran, Iran}
	\author{Shant Baghram}
	\email{baghram@sharif.edu}
	\address{Department of Physics, Sharif University of
		Technology, P.~O.~Box 11155-9161, Tehran, Iran}
		\address{Research Center for High Energy Physics, Department of Physics, Sharif University of Technology, Tehran 11155-9161, Iran}


\begin{abstract}
	The second data release of the Dark Energy Spectroscopic Instrument (DESI), based on more
than fourteen million galaxies and quasars, prefers a dark energy component whose equation
of state is dynamical. In the simplest two-parameter description of this behavior, the
Chevallier--Polarski--Linder (CPL) parametrization, the data favor the quadrant
$w_0>-1$, $w_a<0$, which implies that the equation of state crosses the phantom divide
$w=-1$ at a redshift $z\simeq0.4$.  We show that the local limit of nonlocal gravity, a teleparallel extension of general relativity governed by a scalar function called	gravitational susceptibility $S(x)$, offers a natural explanation. In this model a	cosmological constant is not allowed; dark energy is necessarily dynamical. The phantom crossing seen by an	observer is an effective phenomenon generated by the time evolution of $S(z)$. We	establish a correspondence: to every general-relativistic dark energy model with equation of state $w(z)$, there corresponds a modified TEGR cosmology with $w=-1$ and a suitable susceptibility $S(z)$ that are indistinguishable at the level of the background expansion. We construct the susceptibility functions that reproduce the CPL behavior favored by DESI data, and we confront the model with the combined
CMB,BAO, and SNe data using Boltzmann solver codes. The recovered susceptibility displays a characteristic low-redshift dip with depth  $\beta = -0.025 \pm 0.007$ at $z\simeq 0.4 $,  and its derivative changes sign at a redshift equal to the phantom-crossing epoch.

\end{abstract}

\maketitle


\section{Introduction}
\label{sec:intro}

The standard cosmological model, spatially flat $\Lambda$CDM, which combines a cosmological constant $\Lambda$ with cold dark matter (CDM), successfully explains a wide range of observations,
most notably the temperature and polarization anisotropies of the cosmic microwave background
(CMB) \citep{Planck:2020npipe}.
Yet the physical nature of dark energy (DE) and dark matter (DM) remains unknown, and
the cosmological constant itself suffers from severe theoretical problems	\citep{Zeldovich:1983cr,Weinberg:1988cp}. It is therefore
not surprising that the search for deviations from cosmological constant solution for an accelerating universe has become one of the main endeavors of observational cosmology today \citep{Peebles:2002gy,Perivolaropoulos:2021jda}.

The Dark Energy Spectroscopic Instrument (DESI) was designed precisely for this purpose
\citep{DESI:2018ymu}. It measures the baryonic acoustic oscillation (BAO) through the large scale structure survey.
The BAO data alone are described well by a flat $\Lambda$CDM model, but the parameters
preferred by BAO are in mild tension with those determined from the CMB: the
$\Lambda$CDM fit to DESI$+$CMB shows a $2.3\sigma$ tension. This tension disappears
when the dark energy equation of state is allowed to vary with time
\citep{DESI:2025zgx,RoyChoudhury:2025dhe,RoyChoudhury:2025iis}.

The statistical preference for dynamical dark energy, and in particular the exact
significance, depends on the data combination and on the assumed parametrization
\citep{Khosravi:2015boa,Colgain:2024mtg,Rodrigues:2025,Cortes:2025joz,Colgain:2026ryf,Lu:2025gki}. What matters for the present paper is that the data
consistently point toward a dark energy equation of state that is dynamical and crosses
$w=-1$ at low redshift, known as phantom crossing, and that this pattern is difficult to realize in general relativity with a single field \citep{Cai:2009zp,Vikman:2005,Nojiri:2026uvn}. A phantom component is problematic in general relativity (GR) because it
violates the null energy condition \citep{Caldwell:1999ew,Caldwell:2003vq}.  Modified gravity offers a way
out: the crossing may be an effective description of a modified expansion history rather
than a property of the dark energy fluid itself \citep{Bamba:2008hq,Linder:2025pqt,Calderon:2026hbr,Nojiri:2025uew, Nojiri:2025low, Odintsov:2026mjb,Wen:2026xlj}. Various proposals have been put forward
to accommodate the crossing, ranging from multi-field models \citep{Cai:2009zp,Bedroya:2025fwh} to
modified-gravity constructions in which the crossing is only an effective feature
\citep{Efstratiou:2025}.

In this paper we show that the local limit of nonlocal gravity (NLG) provides a well-motivated gravitational
framework in which the phantom crossing appears naturally as an effective
phenomenon. NLG is a classical, nonlocal generalization of GR
inspired by the nonlocal electrodynamics of media \citep{Hehl:2008a,Hehl:2009b,BMB}.
This theory is formulated as a modification in the teleparallel gravity framework, and its
local limit, obtained when the nonlocal kernel collapses to a delta function, resulting  in the
modified teleparallel equivalent of general relativity (modified TEGR, or MTEGR) model that differs from the base TEGR
by a dimensionless scalar function $S(x)$, known as the gravitational susceptibility
\citep{Tabatabaei:2024base}. 

The cosmological consequences of MTEGR were studied extensively in ~\cite{Tabatabaei:2024base}. One of the important results of this study is that out of different standard FLRW models, only the modified Cartesian flat model is consistent with a time-dependent susceptibility function, and in this model de Sitter
spacetime is not a valid solution. 
A cosmological constant is therefore excluded;
instead, dark energy must be dynamical. The modified continuity equation of this model implies that the
energy density of each component scales with the
susceptibility, enforcing dynamics on every component, including dark energy (with, for example, $w=-1$)
\citep{Tabatabaei:2023b,Tabatabaei:2024base}. The model was shown to be capable of
improving the $H_0$ tension \citep{Riess:2019cxk,Riess:2021jsh,Tabatabaei:2024base}, and
its linear perturbations were studied and tested with the CMB data in \cite{Tabatabaei:2023c} for tighter constraints.

An observer living in the MTEGR universe with a constant DE pressure to density ratio (e.g. $w=-1$), who interprets the data with the standard Friedmann equations, may infer the effects of time evolution of $S(z)$ on all densities as a dynamical equation of state for DE. This is the key observation that motivates the present work.
We show that the resulting effective equation of state $\wbar_{\rm
	de}(z)$ can cross $-1$ even though the true dark energy component  has fixed equation of state ($w=-1$) at all
times. The crossing is therefore not a property of the dark energy fluid itself, but a
consequence of the modified gravitational framework. Given any DE model with known equation of state $w(z)$ within
general relativity, we find the susceptibility $S(z)$ that reproduces it at the
background cosmology level. Applying this procedure to the CPL model preferred by DESI DR2, we
obtain the required $S(z)$. The background analysis gives us a baseline ansatz for $S(z)$, that proves to be helpful in a full (including perturbation) cosmological analysis. Using Monte Carlo Markov Chain (MCMC)
analysis, with the CMB, the DESI DR2 BAO data and different supernovae catalogs, we constrain the ansatz parameters and approve that the model is preferred according to the reported Akaike information criterion (AIC), and deviance information criterion (DIC) \citep{Akaike:1974,Spiegelhalter:2002} by the data, over the standard cosmology model.

The structure of this paper is as follows. In Section \ref{sec:DESI}, we review the results of DESI survey. In Section~\ref{sec:model}, we summarize the local limit
of nonlocal gravity.
In Section~\ref{sec:effective} we derive the effective dark energy density, pressure and
equation of state inferred by an observer in this model, and establish the correspondence
between $w(z)$ models and $S(z)$ models. In Section~\ref{sec:ansatz} we set up the
phenomenological ansatz for $S(z)$ guided by the CPL target. Section~\ref{sec:data}
describes the data sets and the numerical tools. Section~\ref{sec:results} contains the
results of the analysis, and Section~\ref{sec:discussion} discusses the physical implications
of the phantom crossing within this framework and we present our conclusion.

\section{DESI DR2 and phantom crossing}
\label{sec:DESI}
The Dark Energy Spectroscopic Instrument (DESI) was designed for the purpose of detecting the deviations of cosmological expansion from the standard $\Lambda$CDM model
\citep{DESI:2018ymu}. Its second data release (DR2)~\citep{DESI:2025zgx} contains spectra of
more than fourteen million galaxies and quasars. From this sample, the collaboration
extracted baryon acoustic oscillation measurements in a number of redshift bins spanning
the range $0.1\lesssim z \lesssim 4.2$, with the highest-redshift bins coming from the
Lyman-$\alpha$ forest~\citep{DESI:2025lya,DESI:2026lnd}. The DR2 BAO data are consistent with the DR1 measurements and with
the earlier SDSS results, and the distance-redshift relation they trace agrees with that
of the recent supernova compilations over the same redshift range~\citep{DESI:2025zgx}.

The combined DESI$+$CMB data exclude $\Lambda$CDM at $3.1\sigma$,
and adding supernovae raises the significance to $2.8$--$4.2\sigma$ depending on the
sample used.

The most striking feature of the DESI DR2 results is that the preferred dynamical dark
energy crosses the phantom divide. The simplest and most widely used description of a
time-varying dark energy equation of state is the phenomenological Chevallier--Polarski--Linder (CPL)
parametrization~\citep{Chevallier:2001qy,Linder:2002et},
\begin{equation}
	\label{eq:cpl}
	w(a)=w_0+w_a(1-a)\,,
\end{equation}
or, equivalently, $w(z)=w_0+w_a z/(1+z)$. For the combination of DESI BAO, CMB and
the DES Year-5 supernovae (DESY5)~\citep{des_sne}, the DESI collaboration reports
~\citep{DESI:2025zgx}
\begin{equation}
	\label{eq:desi-cpl}
	w_0=-0.752\pm0.057\,,\qquad w_a=-0.86^{+0.23}_{-0.20}\,,
\end{equation}
which deviates up to $4.2\sigma$ from base $\Lambda$CDM. 
These numbers are consistent with, and sharper
than, the DR1 findings, for which the corresponding preference was $2.6\sigma$ with CMB
alone and up to $3.9\sigma$ with supernovae \citep{DESI:2024mwx}.

The crucial feature of the result in equation  \eqref{eq:desi-cpl} is the sign pattern
$w_0>-1$, $w_a<0$. Using equations \eqref{eq:cpl} and \eqref{eq:desi-cpl}, one finds that the
equation of state at the present epoch is quintessence-like, $w(z=0)=w_0>-1$, while
$w(z)=w_0+w_a z/(1+z)$ decreases with redshift and crosses the phantom divide, $w=-1$,
at
\begin{equation}
	\label{eq:crossing}
	z_\times=\frac{-(1+w_0)}{1+w_0+w_a}\simeq0.405\,.
\end{equation}
For $z>z_\times$ the preferred dark energy is phantom, $w<-1$. This crossing is one of
the important features of the DESI DR2 data~\citep{DESI:2025zgx}. \\
Alcock-Paczyński Measurements from the Lyman-$\alpha$ Forest in a recent study put tighter constraints on the dynamics of DE in the matter-dominated era as well	\citep{DESI:2026lnd}.

\section{Local limit of nonlocal gravity}
\label{sec:model}
In this section we summarize the fundamentals of the local limit of nonlocal gravity, which is the theoretical framework underlying our analysis. We begin by reviewing how the nonlocal constitutive relation of NLG reduces to a local one governed by a single scalar susceptibility  $S(x)$. We then introduce the modified Cartesian flat cosmology as a solution to the field equations of this model in the case of a time-dependent $S$.
\subsection{From nonlocal gravity to modified TEGR}
Nonlocal gravity is a classical extension of GR constructed in analogy
with the nonlocal electrodynamics of media~\citep{Hehl:2008a,Hehl:2009b,BMB}. The theory is
formulated in the teleparallel gravity framework, in which spacetime is parallelized by a preferred
tetrad frame field $e^\mu{}_{\hat\alpha}(x)$, and the gravitational field
strength is the torsion tensor $C_{\mu\nu\rho}$ and contorsion tensor $	K_{\mu \nu \rho}$ of the Weitzenb\"ock connection, defined as
\begin{align}
	&	C_{\mu \nu}{}^{\alpha}=e^\alpha{}_{\hat{\beta}}\left(\partial_{\mu}e_{\nu}{}^{\hat{\beta}}-\partial_{\nu}e_{\mu}{}^{\hat{\beta}}\right)
	,  \\ 
	&	K_{\mu \nu \rho} = \frac{1}{2}\, (C_{\mu \rho \nu}+C_{\nu \rho \mu}-C_{\mu \nu \rho})\,.
\end{align}

GR, when written in this language, becomes the teleparallel equivalent of general
relativity (TEGR), and the Einstein equation is rewritten in a new form,
\begin{equation}\label{tele-ein}
	\frac{\partial}{\partial x^\nu}\,\mathfrak{H}^{\mu \nu}{}_{\hat{\alpha}}+\frac{\sqrt{-g}}{8\pi G}\,\Lambda\,e^\mu{}_{\hat{\alpha}} =\sqrt{-g}\,(T_{\hat{\alpha}}{}^\mu + \mathbb{T}_{\hat{\alpha}}{}^\mu)\,,
\end{equation}
in which the gravitational field tensor $\mathfrak{H}_{\mu\nu\rho}$ is
related to the auxiliary torsion tensor, $\mathfrak{C}_{\alpha \beta \gamma} :=C_\alpha\, g_{\beta \gamma} - C_\beta \,g_{\alpha \gamma}+K_{\gamma \alpha \beta}$,  by a simple, local constitutive relation
\begin{equation}\label{gr-const}
	\mathfrak{H}_{\mu \nu \rho}:= \frac{\sqrt{-g}}{8\pi G}\,\mathfrak{C}_{\mu \nu \rho}\,. 
\end{equation}
This is in close analogy
with $(\mathbf{D},\mathbf{H})$ versus $(\mathbf{E},\mathbf{B})$ in electrodynamics.
The term $ \mathbb{T}_{\hat{\alpha}}{}^\mu$ in equation ~\eqref{tele-ein} is the energy-momentum corresponding to the gravitational field, defined as
\begin{equation}
	\mathbb{T}_{\mu\nu} = (\sqrt{-g})^{-1} \left(C_{\mu\rho\sigma}\mathfrak{H}_{\nu}{}^{\rho\sigma} - \frac{1}{4}g_{\mu\nu}C_{\rho\sigma\delta}\mathfrak{H}^{\rho\sigma\delta}\right).
\end{equation}	

In NLG, the constitutive relation (equation \eqref{gr-const}) is modified to become nonlocal: the excitation at an event receives
contributions from the past history of the field, weighted by a causal kernel
$\mathcal{K}(x,x')$~\citep{BMB,Mashhoon:2022ynk},
\begin{align}\label{T1}
	&	\mathfrak{H}_{\mu \nu \rho} = \frac{\sqrt{-g}}{8\pi G}(\mathfrak{C}_{\mu \nu \rho}+ N_{\mu \nu \rho})\,, 
	\\ 
	&	N_{\hat \mu \hat \nu \hat \rho}(x) := \int  \mathcal{K}(x, x')\,\mathfrak{C}_{\hat \mu  \hat \nu  \hat \rho }(x') \sqrt{-g(x')}\, d^4x' \,.
\end{align}

Using this ansatz and substituting  $\mathcal{H}_{\mu \nu \rho}$ back into the TEGR field equations (equation \eqref{tele-ein}), we obtain a set of integro-differential equations that can be solved in various simplifying regimes to extract physical features and connect to phenomenology.

The local limit of this nonlocal relation is obtained by
taking the kernel to be proportional to a four-dimensional Dirac delta function,
\begin{equation}
	\mathcal{K}(x,x')=\frac{S(x)}{\sqrt{-g(x)}}\,\delta(x-x')\,,
\end{equation}
where $S(x)$ is a dimensionless scalar function \citep{Tabatabaei:2024base}. In this limit
the constitutive relation becomes
\begin{equation}
	N_{\mu\nu\rho}=S(x)\,\mathfrak{C}_{\mu\nu\rho}(x)\,.
\end{equation}
The resulting theory, modified TEGR
(MTEGR), is TEGR extended by a single scalar function $S(x)$, the
gravitational susceptibility, which is characteristic of the background spacetime in
the same sense that the electric permittivity and magnetic permeability are
characteristic of a medium in electrodynamics. For $S=0$ one recovers TEGR and hence
general relativity. The attractive nature of gravitational force implies $1+S>0$. There is no field equation for
$S(x)$; like the constitutive functions of a medium, it must be determined from observation, see \cite{Tabatabaei:2024base,Mohammadi:2025mjx}.

\subsection{The modified Cartesian flat model}
A spatially flat FLRW universe, whose tetrad frame is Cartesian ($
e_{\mu}{}^{\hat\alpha}=a(\eta)\delta_\mu^\alpha\,,\quad e^{\mu}{}_{\hat\alpha}=a^{-1}(\eta)\delta^\mu_\alpha$), is a solution of the MTEGR field equations that we will use as a basis for our cosmological analysis. For any other FLRW
representation, the field equations force $S$ to be constant.  The Cartesian flat case allows for a time-dependent
susceptibility, which is the natural companion of a dynamical background~\citep{Tabatabaei:2024base,Tabatabaei:2023a}.

The background dynamics of this model is governed by the modified Friedmann equations, derived from the modified field equations of MTEGR.
With cosmic time $t$, scale factor $a(t)$, Hubble parameter $H=\dot a/a$, and the total
energy density and pressure of matter, radiation and dark energy, $\rho=\sum_i\rho_i$ and
$P=\sum_i P_i$, one finds \citep{Tabatabaei:2024base}
\begin{align}
	\label{eq:f1}
	&3(1+S)H^2 = \Lambda + 8\pi G\,\rho\,,\\
	\label{eq:f2}
	&2(1+S)\frac{\ddot a}{a}+(1+S)H^2 = \Lambda -8\pi G\,P -2\frac{dS}{dt}H\,.
\end{align}
For $S=0$, these reduce to the standard Friedmann equations. Two properties of
equations \eqref{eq:f1} and \eqref{eq:f2} are crucial for what follows. First,  de Sitter
spacetime is not a solution of the modified Cartesian flat model: setting $\rho=P=0$ in equations
\eqref{eq:f1} and \eqref{eq:f2} leads to a contradiction unless $dS/dt=0$
\citep{Mashhoon:2022ynk,Tabatabaei:2024base}.  A cosmological constant is
therefore excluded, and dark energy must be a dynamical component. There is more detail on this subject in Appendix \ref{de-sitter}. Second, the
susceptibility enters the dynamics of every component: differentiating equation
\eqref{eq:f1} and using equation \eqref{eq:f2}, one obtains the modified continuity equation ($\Lambda=0$)
\cite{Tabatabaei:2024base}
\begin{equation}
	\label{eq:cont}
	\frac{d\rho_i}{dt}+3H(\rho_i+P_i)+\frac{\dot S}{1+S}\,\rho_i=0\,,
\end{equation}
which is assumed to hold separately for each component $i$ with $P_i=w_i\rho_i$. Its solution is
\begin{equation}
	\label{eq:rhoi}
	\rho_i(t)=\rho_i(t_0)\,\frac{1+S(t_0)}{1+S(t)}\,a(t)^{-3(1+w_i)}\,,
\end{equation}
where $t_0$ denotes the present epoch ($a(t_0)=1$). The factor $[1+S(t_0)]/[1+S(t)]$
is the imprint of this modified gravity model on every density. In particular, for a dark energy
component with $w=-1$,
\begin{equation}
	\label{eq:rhode}
	\rho_{\rm de}(t)=\rho_{\rm de}(t_0)\,\frac{1+S(t_0)}{1+S(t)}\,,
\end{equation}
so that even the ``cosmological-constant-like'' component is dynamical: its density
follows the inverse of $1+S(t)$ and is constant only if $S$ is constant
\citep{Tabatabaei:2024base,Tabatabaei:2023b}. This is the origin of dynamical dark energy in MTEGR, but it does not yet imply a dynamical equation of state. To achieve that and allow for phantom crossing, we will adopt a different approach in subsequent chapters.

\subsection{Cosmological perturbations}
\label{sec:pert}

A full confrontation of the model with the CMB data requires the linear
perturbation equations about the modified Cartesian flat background. Here we collect the
essential ingredients, and refer the reader to Reference~\cite{Tabatabaei:2023c} for the
complete derivation and for the synchronous-gauge equations that are actually implemented
in the Boltzmann code. Because MTEGR is a tetrad theory, one perturbs the frame field
rather than the metric,
\begin{align}
	\label{eq:pert-tetrad}\nonumber
	e_{\mu}{}^{\hat\alpha}(x)&=a(\eta)\bigl[\delta_\mu^\alpha+\psi_\mu{}^{\hat\alpha}(x)\bigr]\,,\\
	e^{\mu}{}_{\hat\alpha}(x)&=a^{-1}(\eta)\bigl[\delta^\mu_\alpha-\psi^\mu{}_{\hat\alpha}(x)\bigr]\,,
\end{align}
where $\eta$ is conformal time and $\psi_\mu{}^{\hat\alpha}$ is treated to
linear order away from the Minkowski background. The susceptibility is likewise split into
a time-dependant background and an inhomogeneous perturbation,
\begin{equation}
	\label{eq:pert-S}
	S(\eta,\mathbf{x})=\bar S(\eta)+\delta S(\eta,\mathbf{x})\,.
\end{equation}

Adopting the conformal Newtonian gauge and the standard scalar--vector--tensor
decomposition of the tetrad perturbation \citep{Golovnev:2018wbh,Hohmann:2020vcv}, the scalar sector is described by the three
gravitational potentials $\phi$, $\zeta$, and $\psi$ together with $\delta S$. These scalar fields are related to the tetrad perturbation via the relation
\begin{equation}\label{scalar-pert}
	\psi_{00} = \psi\,, \qquad \psi_{0i} = -\psi_{i0} = \partial_i \zeta\,, \qquad  	\psi_{ij} = \delta_{ij}\, \phi\,.
\end{equation}
Using the orthonormality condition of the tetrads, $e_{\mu}{}^{\hat{\alpha}}e_{\nu}{}^{\hat{\beta}} \eta_{\hat{\alpha}\hat{\beta}}=g_{\mu\nu}$, we can connect the metric and  tetrad perturbations together. Here the nonzero components of metric perturbation are
\begin{equation}\label{metric-pert}
	g_{\mu\nu} =- a^2(\eta)(1-2\psi)\,, \quad 	g_{ij} = a^2(\eta) [(1+2\phi)\delta_{ij}]\,.
\end{equation}

The field equations of MTEGR tie these quantities together algebraically,
\begin{equation}
	\label{eq:pert-constraints}
	\phi-\psi+\ss\,\zeta=0\,,\qquad \mathcal{H}\,\delta S= \frac{d\bar{S}}{d\eta}\,\phi\,,
\end{equation}
where $\mathcal{H}=\frac{da}{d\eta}/a$ and we have introduced the shorthand
$\ss\equiv\frac{d\bar{S}}{d\eta}/(1+\bar S)$. The first relation shows that the gravitational slip is
sourced by the susceptibility through the non-dynamical field $\zeta$, while the second
links the perturbation of the medium to the Newtonian potential. The scalar potentials
obey the first-order evolution equation
\begin{equation}
	\label{eq:pert-poisson}
	\frac{d\phi}{d\eta}+\Bigl(-\frac{1}{3\mathcal{H}}\Delta+\mathcal{H}+\frac{1}{2}\ss\Bigr)\phi
	=\frac{\kappa\,a^2}{6(1+\bar S)\mathcal{H}}\,\delta\rho\,.
\end{equation}

The scalar sector is completed by the modified Poisson equation, which
sources the Newtonian potential with the matter inhomogeneity,
\begin{equation}
	\label{eq:pert-poisson2}
	-\Delta\phi + 3\mathcal{H}\bigl(\frac{d\phi}{d\eta}+\mathcal{H}\psi\bigr)
	= \frac{\kappa\,a^2\,\delta\rho - 3\mathcal{H}^2\,\delta S}{2(1+\bar S)}\,,
\end{equation}
which reduces to the familiar $\Delta\phi=4\pi G\,a^2\,\delta\rho$ in the
general-relativistic limit $\ss=0$ \citep{Amendola:2015ksp}.

Every modification of the perturbation equations is controlled by the
single background function $\ss(\eta)$, so that once the ansatz
for $S(z)$ is fixed, the dynamics of the perturbations is fully specified and the model
contains no new free functions or parameters beyond those already present at the
background level. We solve the coupled system above in the synchronous gauge \citep{Ma:1994dv,Lewis:1999bs} inside a
modified version of the Boltzmann code \texttt{CLASS} (Section~\ref{sec:data}), which supplies
the CMB angular power spectra used in our MCMC analysis.
\subsection{Newtonian limit}

In the Newtonian regime, where the susceptibility is constant over the local 
environment ($S(x)=\mathbb{S}$), the modified field equations yield the 
equation
\begin{equation}
	\nabla^2\Phi=\frac{4\pi G\,\rho}{1+\mathbb{S}}\,,
\end{equation}
which is equivalent to Newtonian gravity with $G \to G/(1+\mathbb{S})$. 

Local measurements of the gravitational constant  do 
not constrain $\mathbb{S}$ directly \citep{Tabatabaei:2024base}. Because the bare gravitational constant 
$G$ is a free parameter, experiments measure the degenerate combination
\begin{equation}
	G_{\mathrm{obs}} \equiv \frac{G}{1+\mathbb{S}}\,,
\end{equation}
and any constant value of $\mathbb{S}$ can be absorbed into a redefinition of $G$. 
Thus, as long as the susceptibility is constant locally, solar system physics 
places no bound on its magnitude.

The situation changes when $S(z)$ evolves cosmologically. A time-dependent 
susceptibility introduces $\dot{S}$ terms into the cosmological equations, breaking 
the static degeneracy. The condition $S(0) \simeq 0$ adopted in 
Section~\ref{sec:ansatz} is therefore a normalization choice: it fixes the bare 
$G$ such that the present-day cosmological gravitational strength matches the 
locally measured $G_{\mathrm{obs}}$. With 
this normalization, the low-redshift variation of $S(z)$ that drives phantom 
crossing is confined to a genuine large-scale cosmological effect, safely 
decoupled from local physics.

We note that in the Newtonian regime of the parent nonlocal gravity theory, the 
nonlocal kernel generates an effective dark matter component that explains galaxy 
rotation curves \citep{Rahvar:2014yta}; in the local limit, this halo collapses 
onto the source (due to local kernel $K(x,x')\sim\delta(x-x')$), so MTEGR itself is not a dark matter substitute at the galactic 
scale \citep{Tabatabaei:2024base}.

\section{Effective dark energy}
\label{sec:effective}
In this section, we use a background-level analysis to establish an effective connection between dynamical dark energy within the standard gravitational framework of cosmology and the cosmology of the MTEGR model, with a specific time evolution of the susceptibility function.

\subsection{The effective dark energy density}

An observer in the MTEGR universe measures distances, redshifts, and the expansion 
rate $H(z)$. All cosmological observables are ultimately functions of $H(z)$ through 
the comoving distance
\begin{equation}
	r(z) = \int_0^z \frac{dz'}{H(z')}\,,
\end{equation}
from which angular diameter distances $D_A(z) = r(z)/(1+z)$ and luminosity distances 
$D_L(z) = (1+z)r(z)$ are derived for SNe Ia and BAO analyses, respectively. 
The comparison between the model and data is therefore made through the modified 
Friedmann equation (equation \eqref{eq:f1}). With $1+z=1/a$ and $S(t)=S(z)$, equation~\eqref{eq:rhoi} 
gives
\begin{equation}
	\rho_i(z)=\rho_{i0}\,\frac{1+S(0)}{1+S(z)}\,(1+z)^{3(1+w_i)}\,,
\end{equation}
and equation~\eqref{eq:f1} becomes
\begin{equation}
	\label{eq:fried}
	H^2(z)=\frac{8\pi G}{3}\,\frac{1+S(0)}{\bigl[1+S(z)\bigr]^2}
	\left[(1+z)^3\rho_{m0}+\rho_{\mathrm{de}0}\right],
\end{equation}
where we have neglected relativistic components, which are irrelevant at the redshifts of interest 
here. Defining the critical density by $\rho_c=\rho_{m0}+\rho_{\mathrm{de}0}$, and 
$\Omega_i=\rho_{i0}/\rho_c$, we obtain
\begin{equation}
	\label{eq:H2}
	H^2(z)=H_0^2\left[\frac{1+S(0)}{1+S(z)}\right]^2
	\left[(1+z)^3\Omega_{m0}+\Omega_{\mathrm{de}0}\right],
\end{equation}
where we have used $H_0^2 = 8\pi G \rho_c/[3(1+S(0))]$.

Equation \eqref{eq:H2} is the expansion history of the modified model. Let us now 
ask how this history would be \emph{interpreted} by a standard observer who assumes 
general relativity. Such an observer would parameterize the same $H(z)$ as
\begin{equation}
	\label{eq:gr-form}
	H^2(z) = H_0^2\left[(1+z)^3\Omega_m^{\mathrm{eff}} + \Omega_{\mathrm{de}}^{\mathrm{eff}}(z)\right],
\end{equation}
where $\Omega_m^{\mathrm{eff}}$ is the effective matter density parameter inferred 
from the data under the GR assumption, and $\Omega_{\mathrm{de}}^{\mathrm{eff}}(z)$ 
is the effective dark energy density that would be required to match the observed 
$H(z)$.

Crucially, for an observer analyzing actual data, $\Omega_m^{\mathrm{eff}}$ is not 
a free convention; it is inferred from observables that are sensitive to the 
combination $\Omega_m H_0^2$. The BAO scale provides the strongest such 
constraint. In the GR framework, the sound horizon at the drag epoch,
\begin{equation}
	r_d = \int_{z_d}^\infty \frac{c_s(z)}{H(z)} dz,
\end{equation}
is fixed by the CMB and provides a standard ruler. BAO measurements of the 
transverse and radial scales,
\begin{equation}
	D_A(z)/r_d \quad \text{and} \quad H(z) r_d,
\end{equation}
directly constrain the combination $\Omega_m^{\mathrm{eff}} h^2$ (where 
$h = H_0/(100\,\mathrm{km\,s^{-1}Mpc^{-1}})$) through the shape of the matter 
power spectrum. Similarly, SNe Ia provide a model-independent measurement of 
$D_L(z)$ and therefore of the integrated expansion history, which together with 
BAO breaks degeneracies and yields a best-fit value for $\Omega_m^{\mathrm{eff}}$.

Equating the physical $H(z)$ from the modified theory (equation~\eqref{eq:H2}) with the 
GR-interpreted form (equation~\eqref{eq:gr-form}), we obtain the effective dark energy 
density that the observer would infer
\begin{equation}
	\label{eq:rhobar-de1}
	\Omega_{\mathrm{de}}^{\mathrm{eff}}(z) = - \Omega_m^{\mathrm{eff}}(1+z)^3+ 
	\left[\frac{1+S(0)}{1+S(z)}\right]^2 \left[(1+z)^3\Omega_{m0}+\Omega_{\mathrm{de}0}\right].
\end{equation}

At this point, the physical interpretation of $\Omega_m^{\mathrm{eff}}$ must be 
fixed. For a standard observer, $\Omega_m^{\mathrm{eff}}$ is not arbitrary: it is 
the value that simultaneously fits the BAO power spectrum shape (which constrains 
$\Omega_m^{\mathrm{eff}} h^2$) and the CMB acoustic scale. In the modified theory, 
the true matter density today is $\Omega_{m0}$, and the susceptibility modifies 
the effective gravitational coupling, thereby altering the matter power spectrum 
and the inferred value of $\Omega_m$. A natural and physically motivated choice is
\begin{equation}
	\label{eq:rhom-choice}
	\Omega_m^{\mathrm{eff}} = \frac{\Omega_{m0}}{1+S(0)},
\end{equation}
which ensures that the effective matter density today, as inferred from the 
$z=0$ limit of the Friedmann equation, matches the true density rescaled by the 
present-day susceptibility. This choice is equivalent to identifying the 
combination $8\pi G \rho_{m0}/[1+S(0)]$ as the observed gravitational source 
density, consistent with the Poisson equation in the Newtonian limit. With this 
identification, equation~\eqref{eq:rhobar-de1} becomes
\begin{equation}
	\label{eq:rhobar-de2}
	\Omega_{\mathrm{de}}^{\mathrm{eff}}(z) = 
	\frac{1+S(0)}{\bigl[1+S(z)\bigr]^2}\,\Omega_{\mathrm{de}0} +\left[\left(\frac{1+S(0)}{1+S(z)}\right)^2-1\right](1+z)^3\,\Omega_m^{\mathrm{eff}}.
\end{equation}
At $z=0$, this reduces to
\begin{equation}
	\Omega_{\mathrm{de}}^{\mathrm{eff}}(0) = \frac{\Omega_{\mathrm{de}0}}{1+S(0)},
\end{equation}
which is the natural counterpart of equation~\eqref{eq:rhom-choice}: the effective present 
dark energy density is the true one rescaled by $1+S(0)$.

Equation \eqref{eq:rhobar-de2} is the central object of this section. It shows 
that the dark energy density inferred by an observer, assuming GR, from distance 
measurements (SNe Ia, BAO) differs 
from the true dark energy density of the modified model. The difference is 
controlled entirely by the susceptibility $S(z)$ and by the effective matter 
density $\Omega_m^{\mathrm{eff}}$ that the observer would infer from BAO and 
CMB data. In practice, this means that a time-varying $S(z)$ can produce an 
apparent evolution of the dark energy equation of state, potentially mimicking 
phantom behavior, while the underlying model remains consistent with the measured 
distance-redshift relation and the BAO-standard-ruler constraints.

Importantly, the success of this effective description hinges on whether the 
modified model can simultaneously fit the  primary cosmological probes.
Any deviation from a cosmological constant, including phantom crossing, is then 
an observational signature of the underlying modified gravity theory, encoded in 
$S(z)$ and in the mismatch between the true and inferred matter densities.

\subsection{The effective pressure and the effective equation of state}

The second Friedmann equation (equation~\eqref{eq:f2}) determines the acceleration of the universe,
and with it the effective pressure of dark energy. Setting $P_{\rm de}=-\rho_{\rm de}$
($w=-1$) in equation \eqref{eq:f2}, and expressing the result in terms of redshift through
\begin{equation}
	\frac{d\ln(1+S)}{dt}=-(1+z)H\,\frac{d\ln(1+S)}{dz}\,,
\end{equation}
we find, after using equation \eqref{eq:f1} to eliminate $H^2$,
\begin{align}
	\label{eq:pbar}
	&\Pbar_{\rm de}(z)=
	-\frac{\rho_{{\rm de}0}}{1+S(z)}  -\\ \nonumber
	&\frac{2}{3}\,\frac{d\ln(1+S)}{dz}\,
	\frac{1+S(0)}{\bigl[1+S(z)\bigr]^2}\,(1+z)
	\left[(1+z)^3\rho_{m0}+\rho_{{\rm de}0}\right].
\end{align}
Here $\Pbar_{\rm de}$ is the pressure that an observer using the standard equations would
assign to the  dark energy in order to reproduce the measured acceleration.

The effective equation of state is the ratio of the two,
\begin{equation}
	\label{eq:wbar-def}
	\wbar_{\rm de}(z)=\frac{\Pbar_{\rm de}(z)}{\rhobar_{\rm de}(z)}\,.
\end{equation}
Before evaluating it, it is convenient to introduce a normalized version of the susceptibility. Define
\begin{equation}
	\label{eq:frakS}
	\Sfr(z):=\frac{1+S(z)}{1+S(0)}\,,
\end{equation}
so that $\Sfr(0)=1$. In terms of $\Sfr$ and of the ratio
\begin{equation}
	\chi:=\frac{\rho_{m0}}{\rho_{{\rm de}0}}\,,
\end{equation}
the effective equation of state (equation~\eqref{eq:wbar-def}) takes the compact form
\begin{equation}
	\label{eq:wbar}
	\wbar_{\rm de}(z)= 
	\frac{-\Sfr(z)-\dfrac{2}{3}\,\dfrac{d\ln\Sfr}{dz}\,(1+z)
		\left[(1+z)^3\chi+1\right]}
	{1+\bigl[1-\Sfr^2(z)\bigr](1+z)^3\,\chi}\,.
\end{equation}
This is the main result of this section. Its structure is simple and revealing: If $S$ is constant, then $\Sfr=1$ and equation \eqref{eq:wbar} reduces to
$\wbar_{\rm de}=-1$: the effective dark energy is a cosmological constant, as expected. The deviation from $-1$ is controlled by two terms: the algebraic term
$1-\Sfr^2$ in the denominator,  and the
derivative term $d\ln\Sfr/dz$ in the numerator, which involves the rate of change of the
susceptibility.
The derivative term in the numerator carries a minus sign with respect to
$d\ln\Sfr/dz$: a decreasing susceptibility ($d\ln\Sfr/dz<0$) makes the numerator
less negative and pushes $\wbar_{\rm de}$  above $-1$ (the quintessence side),
while an  increasing susceptibility ($d\ln\Sfr/dz>0$) makes it more negative and
pushes $\wbar_{\rm de}$ {\it below} $-1$, toward the phantom region.

The phantom crossing therefore occurs near the susceptibility's minimum,
meaning when $d\ln\Sfr/dz$ changes sign from negative to positive, provided the terms are
large enough. In the dip picture of Section~\ref{sec:ansatz}, the crossing sits in the
vicinity of the bottom of the dip, on its rising side: as $z$ increases, $\Sfr$ first
decreases (driving $\wbar_{\rm de}$ above $-1$), reaches its minimum, and then increases
again (driving $\wbar_{\rm de}$ below $-1$). This is the mechanism we will exploit.

\subsection{Correspondence between $w(z)$ models and $S(z)$ models}

Equation \eqref{eq:wbar} establishes a  correspondence at the level of the
background, therefore we claim that for every cosmological model based on general relativity in which
dark energy has an equation of state $w(z)$, there exists a cosmology based on MTEGR,
with a suitable susceptibility $S(z)$ and a dark energy component with $w=-1$, such that
the two models are indistinguishable at the level of the background expansion.

\medskip
\noindent The claim follows directly from equation \eqref{eq:wbar}: demanding
$\wbar_{\rm de}(z)=w(z)$ turns this equation into a first-order ordinary differential
equation for $\Sfr(z)$,
\begin{equation}
	\label{eq:ode}
	w(z)=
	\frac{-\Sfr-\dfrac{2}{3}\,\dfrac{d\ln\Sfr}{dz}\,(1+z)
		\left[(1+z)^3\chi+1\right]}
	{1+\bigl[1-\Sfr^2\bigr](1+z)^3\,\chi}\,,
\end{equation}
which can be solved numerically from $z=0$ with the initial condition $\Sfr(0)=1$, for any
prescribed $w(z)$ and any fixed $\chi>0$. The resulting $\Sfr(z)$ reproduces the given
$w(z)$ by construction. For a fixed $\chi$ value (fixed, for example, by the local matter abundance) the mapping is deterministic. Equation \eqref{eq:ode}
has no closed-form solution in general, but it is trivially integrated numerically. We
emphasize that the dark energy component of the modified model never leaves
$w=-1$; the dynamics of the effective equation of state is entirely a gravitational
effect, encoded in $\Sfr(z)$.

This correspondence is the theoretical backbone of the present paper. It tells us that
the DESI DR2 preference for a crossing CPL equation of state does not, by itself,
require phantom matter or any exotic fluid: it can be interpreted as a measurement of
the susceptibility function $S(z)$ of the local limit of nonlocal gravity.

\section{Dynamic dark energy ansatz}
\label{sec:ansatz}
In this section, we take a first step toward the numerical analysis by choosing an appropriate ansatz for $S(z)$ based on the evidence presented thus far.
\subsection{The CPL target}

\begin{figure}
	\centering
	\includegraphics[width=1.0\columnwidth]{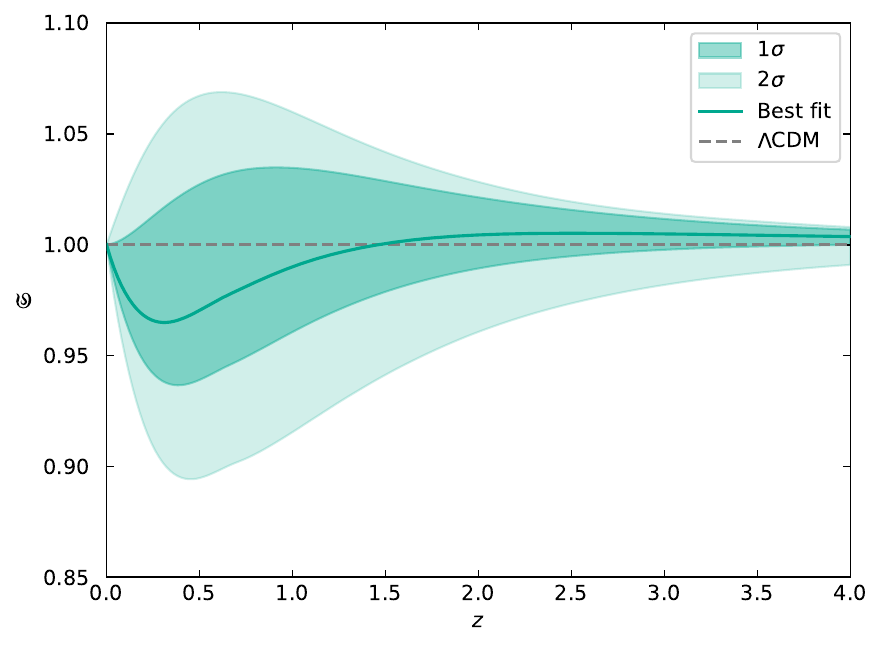}
	\caption{The susceptibility $\Sfr(z)$ from equation~\eqref{eq:ode} for CPL parameters (equation~\eqref{eq:desi-cpl}) with $\chi=0.45$. The solid line is the CMB+BAO+DESY5 best fit; the colored regions are the $1\sigma$ and $2\sigma$ confidence levels of $(w_0,w_a)$. The DESI DR2 sample was used for the plot.}
	\label{fig:plot3-1}
\end{figure}

Guided by the DESI DR2 analysis, which focuses on the CPL parametrization, we now use
equation \eqref{eq:ode} with
\begin{equation}
	w(z)=w_0+w_a\,\frac{z}{1+z}\,,
\end{equation}
and with the best-fit values of equation~\eqref{eq:desi-cpl}, namely
$w_0=-0.752$, $w_a=-0.86$. We fix $\chi=0.45$, consistent with the standard
cosmological determination of the matter-to-dark-energy ratio, and integrate equation
\eqref{eq:ode} from $z=0$ with $\Sfr(0)=1$ toward large redshifts. The result is shown in ~ Fig. \ref{fig:plot3-1}. 
The figure displays the distribution of $\Sfr(z)$ curves obtained by sampling the CPL parameters $(w_0,w_a)$ within their $1\sigma$ and $2\sigma$ confidence regions, while the curve computed with the exact best-fit values in equation \eqref{eq:desi-cpl} is highlighted 	as a solid line. The spread of this family shows how the observational uncertainty on the equation of state translates into an uncertainty in the parameters controlling the shape of $\mathfrak{S}$.

Two lessons follow from this exercise. First, the sign change of $\Sfr'(z)$ and hence
of $S'(z)$ occurs at a redshift near the phantom-crossing redshift of the
CPL model. Second, the dip in $\Sfr(z)$ is what turns, through the derivative term in equation
\eqref{eq:wbar}, into the phantom crossing of the effective equation of state. This is
consistent with the analytic structure of equation \eqref{eq:wbar} discussed above. Although this
background-level reconstruction is not by itself a full cosmological analysis, it tells
us what the susceptibility must look like in order to explain the DESI DR2 data: it
must have a well-defined dip in the low-redshift region $z\lesssim1$, and it must return
to a constant (in practice, to $S(0)$) at higher redshifts, so that the early universe is
unaffected.

\subsection{Choosing the functional form of $S(z)$}

The reconstruction above suggests a phenomenological ansatz for $S(z)$ with the following
properties: (i) a dip at $z\lesssim1$, (ii) recovery to a constant at large $z$, and
(iii) as few parameters as possible. A further simplification is available: the analysis
of the modified model reveals a degeneracy between $H_0$ and $1+S(0)$ (or as previously discussed between $G$ and $1+S(0)$), the two quantities
always appear in combination, so that we may fix $S(0)=0$ without loss of generality.
With this choice, $\Sfr(z)=1+S(z)$.

Our ansatz is a single, fully analytic function, that begins with an additional
simplifying assumption: $\Sfr(\infty)=\Sfr(0)=1$, which reduces the generality here, but many unreleased numerical experiments strongly support it. 
We therefore propose
\begin{equation}
	\label{eq:ansatz-exp}
	\Sfr(z)=1+\bigl(\alpha z+\beta z^2\bigr)e^{-z/\delta}\,.
\end{equation}

This form is continuous and differentiable to all orders, contains the same number of additional parameters as the CPL model, and automatically satisfies $\Sfr(0)=\Sfr(\infty)=1$. In practice,
the parameter $\alpha$ turns out to be strongly constrained to values very close to zero
by the data, so the effective form used in the final analysis, after a reparametrization, and by setting $\alpha=0$, is the two-parameter
function
\begin{equation}\label{exp-form}
	\mathfrak{S}(z)=1 + \frac{\beta}{\delta^2}z^2 e^{2(1-\frac{z}{\delta})}\,.
\end{equation}

This form ensures that the parameter $\delta$ independently determines the dip's position, and $\beta$ determines the dip's height. 
Note that with $S(0)=0$ and $\Sfr=1+S$, the condition $1+S>0$ requires
$\Sfr(z)>0$ for all $z$, which is easily satisfied by this ansatz if $\beta>-1$. It is also easy to see that	$\delta>0$ is a reasonable prior choice, and that a dip would only occur if $\beta<0$. Although we do not enforce this negativity as a prior on $\beta$ in our numerical analysis, the resulting posterior for $\beta$ nevertheless satisfies $\beta<0$. It is also evident that, for small $|\beta|$, the parameter $\delta$ becomes ineffective and has negligible impact on the physics. Consequently, we expect the posterior distributions of $\delta$ to broaden near the $\beta = 0$ line.

\section{Data and methods}
\label{sec:data}
This section introduces and reviews the tools, numerical methods, and data sets used in the analysis.
\subsection{Data sets}
The analysis is based on different data-set permutations that, according to the DESI DR2
collaboration, most strongly support dynamical dark energy
\citep{DESI:2025zgx}. Each of these sets plays a special role in constraining the 
cosmological parameters, and we explore multiple supernova catalogs to assess the 
robustness of our results.

\begin{itemize}
	\item \textbf{CMB.}
	We use the final Planck data release (PR4), based on the NPIPE processing pipeline
	\citep{Planck:2020npipe}. Specifically, we use the full temperature and polarization
	likelihood, including the $TT$, $TE$, and $EE$ power spectra and their covariance
	matrices. We refer to this combination collectively as $TTTEEE$.
	
	\item \textbf{BAO.}
	We use the thirteen BAO measurements from the second DESI data release (DR2)
	\citep{DESI:2025zgx}, covering galaxy, quasar, and Lyman-$\alpha$ tracers.
	These constitute the most precise BAO measurements available to date.
	
	\item \textbf{Supernovae.}
	To test the sensitivity of our results to the choice of supernova catalogue, we perform
	the analysis using three distinct samples: the Dark Energy Survey Year-5 (DESY5) sample
	\citep{des_sne}, the Pantheon+ compilation \citep{Scolnic:2021amg}, and the Union3 sample
	\citep{Rubin:2023rnp}.  As reported by the DESI
	collaboration, the combination involving the DESY5 sample yields the strongest preference
	for a dynamical dark-energy interpretation \citep{DESI:2025zgx}. Thus, we focus on this catalogue for comparing different models. By comparing all three
	configurations (DESY5, Pantheon+, and Union3) we can assess whether any
	inferred deviation from $\Lambda$CDM is driven by the choice of supernova catalogue or
	represents a robust feature of the combined data.
	
\end{itemize}

\subsection{Numerical tools}

As the theoretical platform we use the Boltzmann code \texttt{CLASS}
\citep{CLASS}, which we have modified to implement the modified Cartesian flat model of
MTEGR. This includes modifications in both the background equations \eqref{eq:f1} and \eqref{eq:f2} together with the modified continuity equation in equation \eqref{eq:cont} and the linear perturbation equations of Reference ~\cite{Tabatabaei:2023c}. The modified code is publicly available\footnote{https://github.com/smjty25/class\_MTEGR}. For the parameter estimation we
use the Monte Carlo Markov Chain sampler \texttt{Cobaya} \citep{Cobaya2021}, which
interfaces directly with the modified \texttt{CLASS}. The \texttt{GetDist} \citep{Lewis:2019xzd} code has been used to analyze the chains and to draw triangle posterior plots.

In the code pipeline, the susceptibility function $S(z)$, and its derivatives, are treated as input
functions of the scale factor, and the modified continuity equation (equation~\eqref{eq:cont}) fixes the evolution of all components. The linear perturbation equations of the modified
Cartesian flat model were derived and validated against the CMB in
Reference~\cite{Tabatabaei:2023c}; we employ them here without modification, in the synchronous gauge.

The three models we are going to analyze and compare are CPL, the minimal dynamic dark energy model, EXP, short for the exponential ansatz in equation~\eqref{exp-form}, and also the standard $\Lambda$CDM model. The CPL and EXP models both have two additional parameters compared to the base $\Lambda$CDM model.
\begin{table*}
	\centering
	\begin{tabular}{l|ccc}
		\hline
		Parameter & $\Lambda$CDM  & CPL & EXP  \\
		\hline
		$\Omega_c h^2$ & $0.118 \pm 0.001$ & $0.119 \pm 0.001$ & $0.118 \pm 0.001$ \\
		$\Omega_b h^2$ & $0.0220 \pm 0.0001$ & $0.0220 \pm 0.0001$ & $0.0220 \pm 0.0001$ \\
		$H_0$ & $68.424^{+0.305}_{-0.308}$ & $66.760^{+0.591}_{-0.576}$ & $67.096^{+0.473}_{-0.486}$ \\
		$\ln(10^{10} A_s)$ & $3.067^{+0.043}_{-0.044}$ & $3.056^{+0.047}_{-0.048}$ & $3.063^{+0.042}_{-0.043}$ \\
		$n_s$ & $0.968 \pm 0.003$ & $0.967 \pm 0.004$ & $0.969^{+0.003}_{-0.004}$ \\
		$\tau$ & $0.072 \pm 0.022$ & $0.066^{+0.024}_{-0.025}$ & $0.071 \pm 0.022$ \\
		$w_0$ & --- & $-0.793^{+0.059}_{-0.058}$ & --- \\
		$w_a$ & --- & $-0.594^{+0.233}_{-0.236}$ & --- \\
		$\beta$ & --- &  --- & $-0.025 \pm 0.007$ \\
		$\delta$ & --- & --- & $0.391 \pm 0.114$ \\
		\hline
		$\Delta\chi^2$ &--- & $17.82$ & $14.65$ \\
		$\Delta$AIC & --- & $-13.82$ & $-10.65$ \\
		$\Delta$DIC & --- & $-13.94$ & $-16.43$ \\
		\hline
	\end{tabular}
	\caption{Parameter constraints and model comparison statistics for the CMB+BAO+DESY5 data set.}
	\label{tab:models-desy5}
\end{table*}
\section{Results}
\label{sec:results}

We now report the results of the MCMC analysis of the CMB$+$BAO$+$different supernovae catalogs data with the
MTEGR model.

Our experiments with multiple functional forms for $\mathfrak{S}(z)$, always
select susceptibility functions with the same qualitative shape: $\Sfr(z)$ starts from
$1$, develops a dip of depth controlled by $\beta$ in the low-redshift region, and returns
to $1$ at higher redshifts on a scale set by $\delta$. In most cases the recovered
$\Sfr(z)$ has a minimum at a redshift consistent with the phantom-crossing redshift
$z_\times\simeq0.4$ of the CPL model, and the derivative $d\ln\Sfr/dz$ changes sign
there. Through equation~\eqref{eq:wbar}, this dip translates into an effective dark energy
equation of state that crosses $-1$ from above at $z\simeq z_\times$, precisely the
behavior favored by DESI DR2. In other words, the data, when interpreted within the local
limit of nonlocal gravity, consistently select a susceptibility with a low-redshift dip,
and the dip is the mechanism that generates the phantom crossing.

\begin{figure*}
	\centering
	\includegraphics[width=1.3\columnwidth]{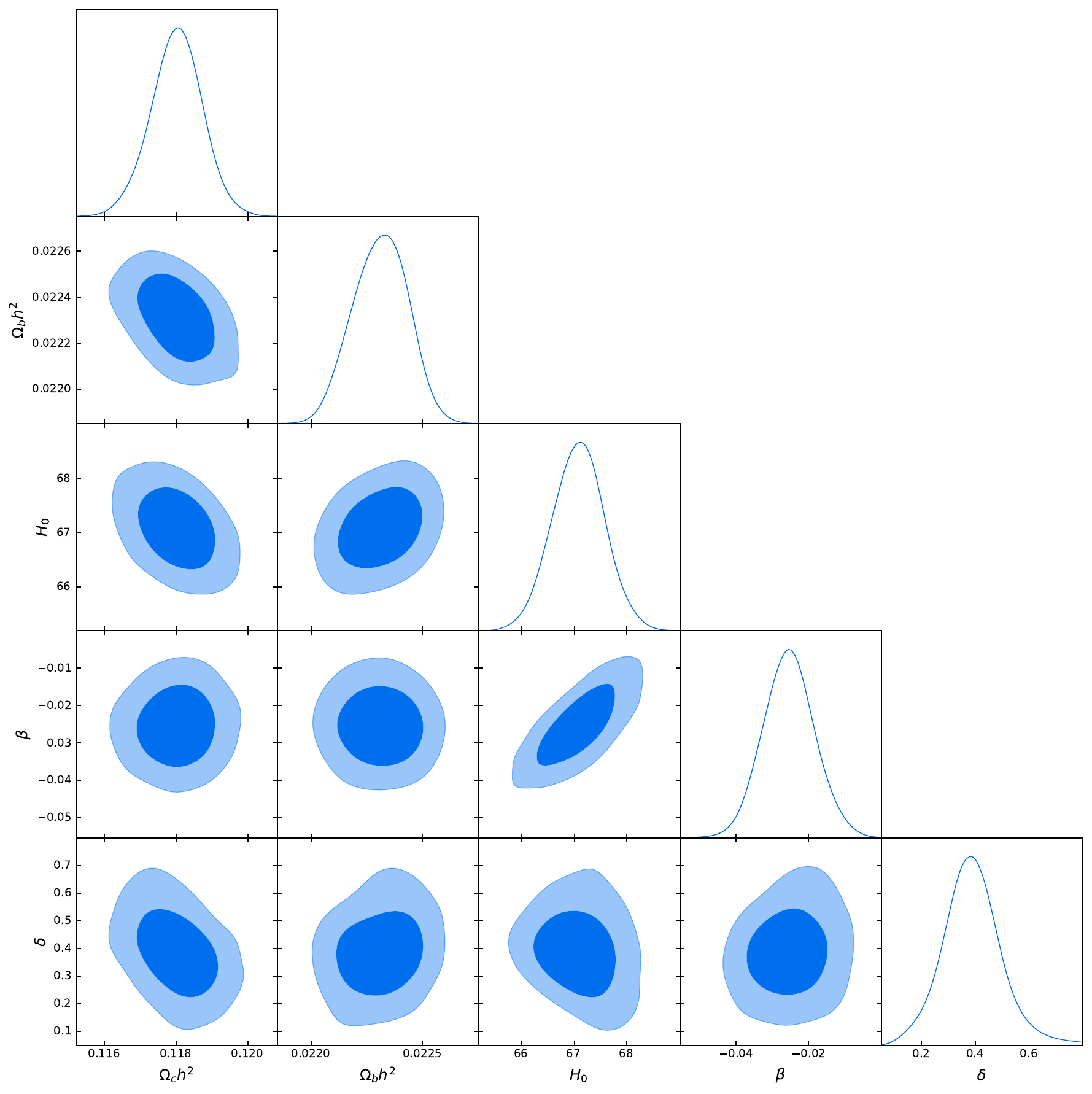}
	\caption{Posterior distributions for the CMB$+$BAO$+$DESY5 data with the exponential
		ansatz $\mathfrak{S}(z)=1 + \frac{\beta}{\delta^2}z^2 e^{2(1-\frac{z}{\delta})}$.}
	\label{fig:plot3-4}
\end{figure*}

Fig.~\ref{fig:plot3-4} presents our main result: a triangle plot of the posterior
distributions obtained from the Markov chains generated for the EXP model using the
CMB$+$BAO$+$DESY5 data combination. The figure shows the credible regions for the
cosmological parameters and the EXP model parameters $\beta$ and $\delta$. We find that
$\delta$ places the dip in the $\Sfr(z)$ function near
$z_\times \sim 0.4$. On the other hand, the best-fit value of $\beta$, together with its
uncertainty, lies more than $2\sigma$ away from the $\Lambda$CDM value,
$\beta=0$. A calculation of the AIC and DIC model-selection criteria, presented in
Table~\ref{tab:models-desy5}, shows that the EXP ansatz is strongly preferred
over the baseline $\Lambda$CDM model according to both criteria. Compared with
the CPL model, however, the EXP ansatz is favored by one criterion and disfavored
by the other, indicating that the two models remain competitive.

\begin{figure*}
	\centering
	\includegraphics[width=1.3\columnwidth]{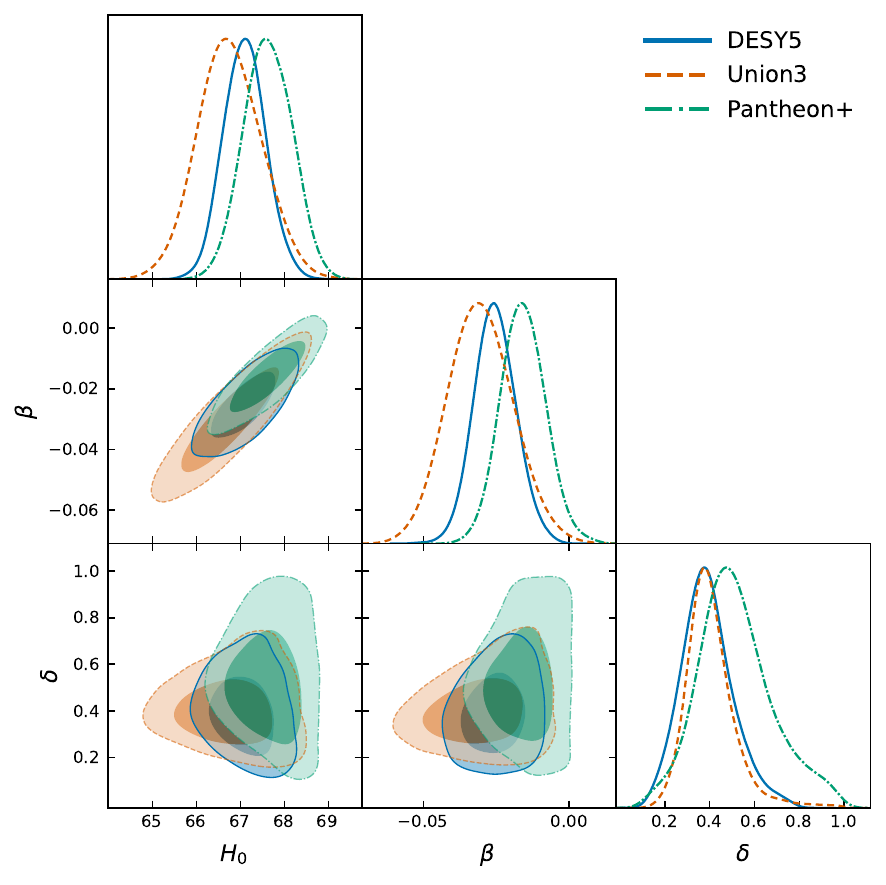}
	\caption{Complete triangle plot of the joint $H_0$--$\beta$--$\delta$
		posterior distributions for the exponential model, comparing the three supernova
		catalogues DESY5, Union3 and Pantheon+.}
	\label{fig:sne-triangle}
\end{figure*}

An important feature of the final posteriors
(Fig.~\ref{fig:plot3-4} and \ref{fig:sne-triangle})  is the  correlation between the depth parameter $\beta$
and the Hubble constant $H_0$. This degeneracy follows directly from the background
relations of Section~\ref{sec:effective}. 
Equation~\eqref{eq:H2} reads
$H^2(z)=H_0^2\,\Sfr(z)^{-2}\,[(1+z)^3\Omega_{m0}+\Omega_{\mathrm{de}0}]$, so that for
a fixed $H_0$ a more negative $\beta$ (a deeper dip in $\Sfr(z)$) enhances the
expansion rate at low redshift, since the factor $\Sfr(z)^{-2}$ grows above unity
where the dip sits. The CMB, however, fixes the angular size of the sound horizon,
$\theta_\ast=r_s/D_A(z_\ast)$, with
$D_A(z_\ast)\propto(1+z_\ast)^{-1}\int_0^{z_\ast}dz/H(z)$. Raising $H(z)$ at low $z$
shortens this comoving distance and therefore enlarges $\theta_\ast$; restoring the
observed value requires lowering $H_0$. The two parameters are thus positively
correlated, which explains both the
tilt of the $H_0$--$\beta$ contour in Fig.~\ref{fig:plot3-4} and \ref{fig:sne-triangle} and the systematic shift of
$H_0$ among the models in Table~\ref{tab:models-desy5}.

\begin{figure*}
	\centering
	\includegraphics[width=1.3\columnwidth]{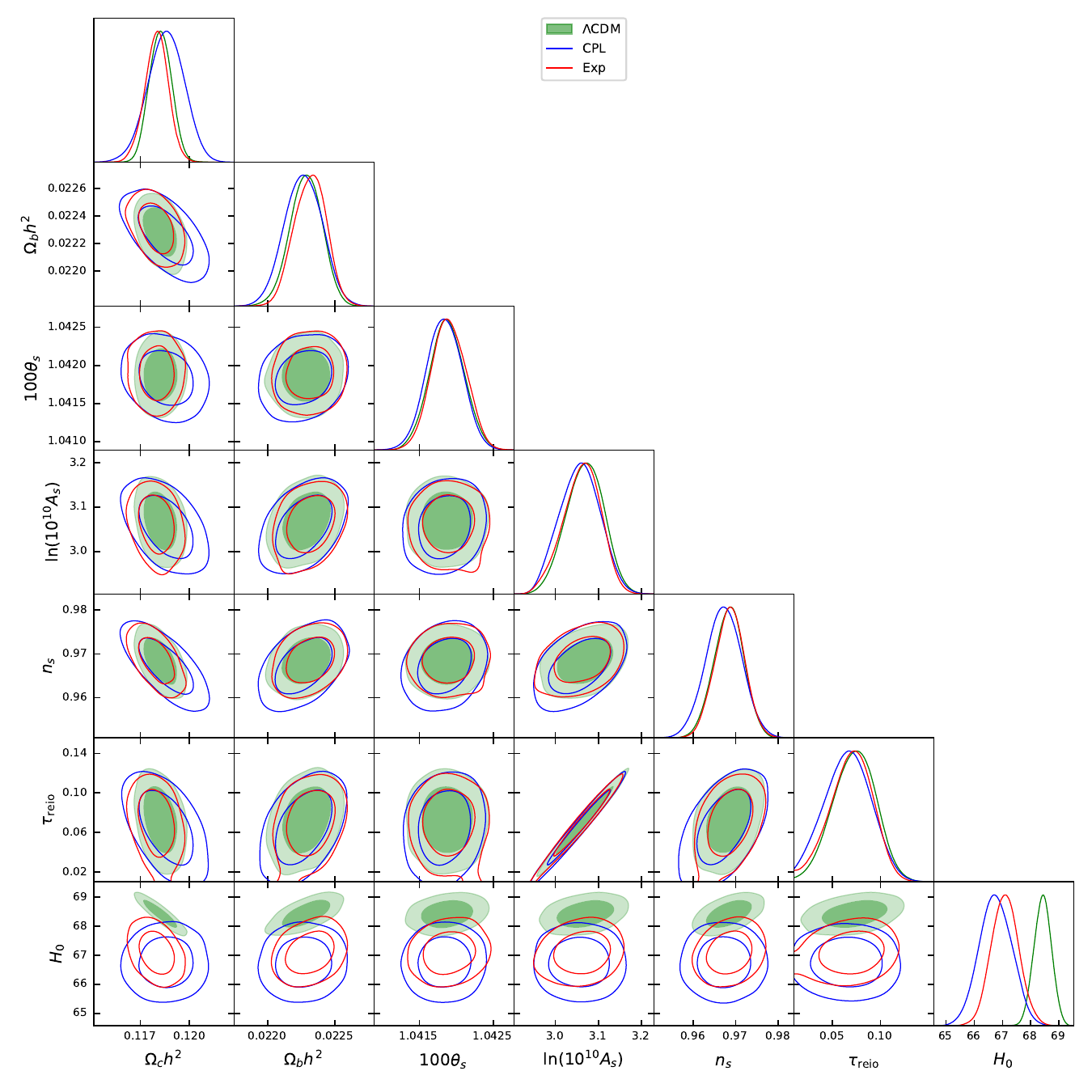}
	\caption{Complete triangle plot comparing the posterior distributions
		obtained for the three models $\Lambda$CDM, CPL and EXP for the CMB+BAO+DESY5 data set.}
	\label{fig:models-triangle}
\end{figure*}

The comparison between the results of different models is displayed in Fig.~\ref{fig:models-triangle}, where the constraints on
the parameters shared by the three models are overlaid. The figure makes it explicit that
$\Lambda$CDM, CPL and EXP differ mainly along the $H_0$ direction, while the common cosmological parameters ($\Omega_ch^2$, $\Omega_bh^2$, $n_s$,
$\ln(10^{10}A_s)$, $	\tau$) remain tightly constrained and essentially model-independent.

\begin{table*}
	\centering
	\begin{tabular}{l|ccc}
		\hline
		Parameter & DESY5 & Pantheon+ & Union3 \\
		\hline
		$H_0$ & $67.096^{+0.473}_{-0.486}$ & $67.623^{+0.565}_{-0.551}$ & $66.756^{+0.738}_{-0.716}$ \\
		$\beta$ & $-0.025 \pm 0.007$ & $-0.016 \pm 0.008$ & $-0.030 \pm 0.011$ \\
		$\delta$ & $0.391 \pm 0.114$ & $0.516^{+0.166}_{-0.158}$ & $0.407^{+0.092}_{-0.098}$ \\
		\hline
	\end{tabular}
	\caption{Parameter constraints for the exponential model using DESY5, Pantheon+, and Union3 supernovae samples.}
	\label{tab:sne-catalogs}
\end{table*}

To verify that our conclusions do not depend on the particular supernova
sample, the EXP MCMC analysis is repeated and reported in Table~\ref{tab:sne-catalogs} for the 
Pantheon+ \citep{Scolnic:2021amg} and Union3 \citep{Rubin:2023rnp} catalogues, with the CMB
and BAO data held fixed. All three catalogues return a negative $\beta$ and a dip position
$\delta$ of the same order, resulting in the same qualitative susceptibility, with only mild shifts in $\beta$ and $\delta$: Union3 prefers the deepest dip
($\beta=-0.030\pm0.011$) together with the lowest $H_0$, whereas Pantheon+ prefers the
shallowest dip ($\beta=-0.016\pm0.008$) and the highest $H_0$. This is precisely the
$\beta$--$H_0$ degeneracy described above, and it shows that the low-redshift dip is a
robust feature of the data rather than an artefact of a single supernova compilation. The
full posterior distributions of $(H_0,\beta,\delta)$ for the three catalogues are
displayed in Fig.~\ref{fig:sne-triangle}.

\section{Discussion}
\label{sec:discussion}

The central result of this paper is that the phantom crossing favored by DESI DR2 can be
realized within the local limit of nonlocal gravity without introducing any phantom
component. In the modified model, the dark energy fluid has $w=-1$ at all times and
satisfies the modified continuity equation (equation~\eqref{eq:cont}); the model does not introduce an explicit 
phantom matter degree of freedom. What crosses $-1$ is the
effective equation of state (equation~\eqref{eq:wbar}), a derived quantity that encodes the
difference between the true expansion history of the modified model and the standard
Friedmann interpretation of that history. 	
This interpretation has a number of appealing features. First, it avoids the well-known
instabilities and fine-tuning problems of phantom fields \citep{Cai:2009zp}. Second, it
makes the crossing generic: any susceptibility with a dip produces a crossing,
and the data select a dip.

The physical content of the crossing is the dip in the susceptibility
$S(z)$ at $z\simeq 0.4$, which represents a transient, low-redshift feature
in the gravitational response of spacetime. This behavior is qualitatively
analogous to the frequency-dependent response of a medium in electrodynamics:
near a resonance, the real part of a material response function can exhibit
rapid variation, including a local extremum or, depending on the model,
a change in sign. In a causal medium, the dispersive and absorptive parts
of the response are related by the Kramers--Kronig relations
\citep{LandauLifshitz:ECM,Jackson:1999}. This analogy is intended to be
qualitative; $S(z)$ is a gravitational susceptibility and should not be
identified directly with an electromagnetic dielectric function.

Several other frameworks have been proposed to explain the DESI DR2 phantom crossing. In
scalar-tensor gravity, a reconstruction of the underlying Lagrangian from the DESI DR2
data was performed in Reference~\cite{Efstratiou:2025}, where the crossing is realized through
a non-minimal coupling and a modified effective gravitational constant; in that model the
inferred $H_0$ is also increased relative to $\Lambda$CDM. In quintom models, the
crossing is achieved with two fields, one quintessence-like and one phantom-like
\citep{Cai:2009zp}. In string-inspired settings, self-dual potentials have been proposed
that mimic the DESI DR2 behavior with a single rolling field near a maximum of its
potential \citep{Anchordoqui:2025}. The present framework differs from all of these in an
essential way: no new matter component, no second field, and no phantom field is
introduced. The crossing is a property of the gravitational sector itself, specifically
of the susceptibility $S(z)$, whose role is analogous to that of a constitutive function
of a medium. In this respect the model is closer in spirit to the electrodynamics-of-media
analogy that motivated nonlocal gravity in the first place \citep{Hehl:2008a,Hehl:2009b}.
\section*{Data availability}
We use public data sets in this work. See Section \ref{sec:data}.

\section*{ACKNOWLEDGMENTS} 
We are grateful to Bahram Mashhoon for introducing us to the concept of nonlocal gravity through his academic guidance and insightful lectures.
We would like to thank Abdolali Banihashemi and Nima Khosravi for insightful comments and suggestions.
J.T. would like to thank Saman Moghimi-Araghi and  Department of Physics, Sharif University Computational Physics
Laboratory for their invaluable support and for providing us with computational resources.
The authors used ChatGPT (OpenAI) for assistance with language editing and manuscript presentation.
SB is supported in part by the Abdus Salam International Centre for Theoretical Physics (ICTP) through the Regular Associateship scheme, and by the Sharif University of Technology Office of the Vice President for Research under Grant No.~G4010204.\\ \\

 \bibliographystyle{apsrev4-2}

\bibliography{phantom_MTEGR_refs} 

\appendix

\section{Absence of a de Sitter solution in MTEGR}
\label{de-sitter}
Because the impossibility of de Sitter spacetime is the physical origin of dynamic dark
energy in this framework, it is worth recalling the argument. Setting $\rho=P=0$ in
equations \eqref{eq:f1} and \eqref{eq:f2}, the first equation gives, for an expanding universe,
\begin{equation}
	\frac{\dot a}{a}=\Bigl(\frac{\Lambda}{3}\Bigr)^{1/2}(1+S)^{-1/2}\,,
\end{equation}
Differentiating this relation with respect to $t$ and eliminating
$\ddot a/a$ with the second Friedmann equation, one arrives at
\begin{equation}
	2\,\frac{\dot a}{a}\,\frac{dS}{dt}
	=\Bigl(\frac{\Lambda}{3}\Bigr)^{1/2}(1+S)^{-1/2}\,\frac{dS}{dt}\,.
\end{equation}
For $dS/dt\neq0$, division by $dS/dt$ leaves a relation that contradicts the first
equation. Hence, for a time-dependent susceptibility, no de Sitter solution exists
\citep{Tabatabaei:2024base}. We therefore set $\Lambda=0$ throughout and account for the
accelerated expansion through a dynamical dark energy component with negative pressure,
$P_{\rm de}=w_{\rm de}\,\rho_{\rm de}$ with $w_{\rm de}<0$. In the modified Cartesian
flat model, $w_{\rm de}=-1$ is  allowed (unlike the standard model, where it is
trivially the cosmological constant), and equation~\eqref{eq:rhode} shows that, even then, the
density evolves with $S(t)$.

\section{Conservation of the effective dark energy}
\label{conv-effective}
A natural question at this point is whether the effective dark energy defined by
equation~\eqref{eq:rhobar-de2} and equation~\eqref{eq:pbar} satisfies a conservation equation. It does not.
A direct calculation gives
\begin{equation}
	\label{eq:noncons}
	(1+z)\frac{d\rhobar_{\rm de}}{dz}-3\bigl(\rhobar_{\rm de}+\Pbar_{\rm de}\bigr)
	=3\rho_{{\rm de}0}\,\frac{S(z)-S(0)}{\bigl[1+S(z)\bigr]^2}\,.
\end{equation}
The right-hand side vanishes only if $S(z)=S(0)$, only if the susceptibility is constant.
In general, the effective dark energy is not conserved: the modified continuity equation
(equation~\eqref{eq:cont}) redistributes energy among the components in a way that, from the
standard viewpoint, looks like an injection or removal of energy into the effective dark
energy sector. This is the same phenomenon that, in the modified Cartesian flat model, was
identified with the production of heat (or negative entropy production) associated with
the variation of $S$ \cite{Tabatabaei:2024base}.


\end{document}